\documentclass{article} 
\usepackage{epsfig,amsbsy}
\makeatletter

          \@addtoreset{equation}{section}
       \makeatother

\begin{document}
\title{%
An estimation of the Higgs boson mass in the two loop approximation
in a noncommutative differential geometry 
}
\author{%
  Yoshitaka {\sc Okumura}
  \thanks{e-mail address: okum@isc.chubu.ac.jp}\\
{\it Department of Natural Science, 
Chubu University, Kasugai, 487, Japan}
}
\maketitle
\begin{abstract}
{
In the two loop approximation, 
a renormalization group analysis of the Higgs boson mass is performed 
based on the  condition $g^2=(5/3)g'^2=4\lambda$ for 
SU(2)$_{\hbox{\rm \tiny L}}$, U(1)$_{\hbox{\rm\tiny Y}}$ 
 gauge  and the Higgs quartic coupling constants, respectively.   
This  condition is introduced in the new scheme 
of our noncommutative differential geometry (NCG)
 for the reconstruction of the standard model. 
However, contrary to ${\rm SU(5)}$ GUT without 
supersymmetry, the grand unification of coupling constants is not 
realized in this scheme.
The physical mass of the Higgs boson  
depends strongly 
on the top quark mass $m_{\hbox{\rm\scriptsize top}}$ 
through the Yukawa coupling of
the top quark in the $\beta$ functions.
The two loop effect lowers the numerical value calculated within 
the one loop approximation by several GeV.
The Higgs boson mass varies from 150.25GeV to 167.15GeV
corresponding to $169.34{\rm GeV}\leq 
m_{\hbox{\rm\scriptsize top}}\leq 181.02{\rm GeV}$. 
We find $m_{\hbox{\rm\tiny H}}=158.18$GeV 
for $ m_{\hbox{\rm\scriptsize top}}=175.04$GeV and 
$m_{\hbox{\rm\tiny H}}=166.12$GeV for 
$ m_{\hbox{\rm\scriptsize top}}=180.34$GeV.
}

\vskip 0.2cm
\noindent
\end{abstract}
\thispagestyle{empty}
\section{Introduction}
The Higgs mechanism is essential for any spontaneously broken
gauge theory. Its presence ensures 
the renormalizability of the theory and
makes the theory realistic by giving masses to particles, such as
gauge and matter fields,
through the vacuum expectation value of the Higgs boson field.
The standard model in particle physics also involves the Higgs mechanism
and shows remarkable agreement with existing data. After 
the discovery of the top quark in 1994, the only undetected particle 
in the standard model is the Higgs boson. Now,  studies concerning
the Higgs boson search 
are being conducted in both theoretical and 
experimental settings and it is expected that the Higgs boson
will be discovered within a decade in  future experiments 
at Fermi-Lab and CERN.
\par
Many models (most notably, the
technicolor model, the Kaluza-Klein model and 
recently the approach based on  noncommutative differential geometry (NCG)
on the discrete) have been constructed for the purpose of 
understanding the Higgs mechanism.
Among these, the NCG approach, originally proposed by Connes \cite{Con}, 
provides a unified picture of gauge and Higgs fields as a generalized
connection on the principal bundle with the base space $M_4\times Z_{_N}$.
It should be noted that the NCG approach does not demand any  physical
modes other than the usual one. \par
Many versions of the NCG approach have appeared since Connes's original 
presentation, and the standard model has been successfully reconstructed
using these approaches. The characteristic feature of the reconstruction
of the gauge theory in the NCG approach is to impose  restrictions
on the gauge and the Higgs quartic coupling constants. This is because
the gauge and Higgs fields are represented together 
as a generalized gauge field.
These restrictions yield 
 numerical estimates of the Weinberg angle and 
 mass relations involving the gauge boson and other particles,
such as the Higgs boson  and top quark in tree level.
Several works have appeared \cite{AGM,SS,RGNCG} 
 estimating the quantum effects of these 
relations by assuming them to hold at some renormalization point. 
\par
The present author   
has also proposed an unique formulation based on a NCG \cite{NCGNP,FSTM}.
Our formulation using a NCG employs a  generalization of 
the usual differential geometry on an ordinary manifold to the discrete
manifold $M_4\times Z_{_N}$.
The reconstruction of SO(10) GUT  and the left-right symmetric
gauge model  \cite{GUT} had already been  performed using 
our NCG scheme.
In a NCG on $M_4\times Z_2$,
the extra differential one-form $\chi$ in $Z_2$
is introduced in addition to
the usual one-form $dx^\mu$ in $M_4$, 
and therefore our formulation is very similar
to ordinary differential geometry, 
 contrastingly, in Connes' original scheme 
  the Dirac matrices $\gamma_\mu$ and $\gamma_5$ 
are used  to describe the generalized gauge field.
In a NCG, the  gauge field and the Higgs boson field are given as coefficients
of $dx^\mu$ and $\chi$, respectively, 
in the generalized connection on $M_4\times Z_2$. 
However, there is no symmetry to mix $dx^\mu$ and $\chi$, 
and, therefore, the ordinary gauge field cannot be transformed into 
the Higgs boson field.
In Ref.\ \cite{FSTM}, 
the reconstruction of the standard model is successfully carried
out based on a new scheme of our NCG. Three generations of fermions,
 including 
left and right-handed quarks and leptons, are incorporated. 
In addition, the strong interaction is nicely included in this scheme.
The relations 
$g^2=(5/3)g'^2=4\lambda$ 
are introduced in Ref.\ \cite{FSTM},
where $g, g'$ and $\lambda$ are
SU(2)$_{\hbox{\rm\tiny L}}$ and U(1)$_{\hbox{\rm\tiny Y}}$ 
gauge coupling constants and 
the Higgs quartic coupling constant, respectively.   
However, the grand unification for gauge coupling constants 
is not achieved in this scheme.
The former part of this relation leads
 to $\sin^2\theta_{\hbox{\rm\tiny W}}=3/8$, 
and the latter part leads to the mass relation 
$m_{\hbox{\rm\tiny H}}=\sqrt{2}m_{\mbox{\rm\tiny W}}$.
We assume that the relations $g^2=(5/3)g'^2=4\lambda$ 
hold at one renormalization point.
With this assumption, we can perform the renormalization
group analysis of the running coupling constants $g$, $g'$ and 
$\lambda$, and 
the physical Higgs boson mass can thus be estimated.  
Within the one loop approximation, 
this analysis was carried out in Ref.\ \cite{RGNCG}.
In that context, 
we found $m_{\hbox{\rm\tiny H}}=164.01$GeV for $ m_{\rm top}=175$GeV. 
In this article, we will perform the two loop analysis of the Higgs boson mass
using the same method.\par
\section{ The review of our NCG on $M_4\times Z_2$}
We first briefly review our previous formulation \cite{FSTM},
because it is not well-known among particle physicists.
For a detailed description, we refer the reader to Ref.\ \cite{FSTM}.
\par
Let us start with the equation of the generalized gauge field 
${\cal A}(x,y)$ on the principal bundle with the base space
$M_4\times Z_2$,
\begin{equation}
      {\cal A}(x,y)=\sum_{i}a^\dagger_{i}(x,y)
      {\boldsymbol d}a_i(x,y)+
      \sum_{j}b^\dagger_{j}(x,y)
      {\boldsymbol d}b_j(x,y),\label{2.3}
\end{equation}
where 
$a_i(x,y)$ and $b_j(x,y)$ are  square-matrix-valued functions
and  are taken so as to commute with each other, because
$\sum_{i}a^\dagger_{i}(x,y){\mbox{\boldmath $d$}}a_i(x,y)$ is 
the flavor sector, including the flavor gauge and the Higgs fields, 
while $\sum_{j}b^\dagger_{j}(x,y){\mbox{\boldmath $d$}}b_j(x,y)$ 
corresponds to the color sector. 
The indices $i$ and $j$ are  variables of the extra
internal space which we cannot  identify. 
The operator ${\boldsymbol d}$ in Eq. (\ref{2.3}) 
is the generalized exterior derivative defined as follows:
\begin{eqnarray} 
&&       {\boldsymbol d}=d + d_\chi , \label{2.4}\\  
&&     da_i(x,y) = \partial_\mu a_i(x,y)dx^\mu,\hskip 1cm \label{2.5}\\
&&   d_{\chi} a_i(x,y) =[-a_i(x,y)M(y) + M(y)a_i(x,-y)]\chi,
        \label{2.6}\\
&&      {d}b_j(x,y)= \partial_\mu b_j(x,y)dx^\mu, \label{2.8}\\
&&   d_{\chi} b_j(x,y) =0. \label{2.9}
\end{eqnarray}
Here, 
$dx^\mu$ is the
ordinary one-form basis taken to be dimensionless in Minkowski space 
$M_4$, and $\chi$ 
is the one-form basis also  assumed to be dimensionless 
in the discrete space $Z_2$.
We have introduced the $x$-independent matrix $M(y)$ 
whose hermitian conjugation is given by $M(y)^\dagger=M(-y)$. 
The matrix $M(y)$ determines the scale and pattern of 
the spontaneous breakdown of the gauge symmetry. Thus, 
Eq.(\ref{2.9}) implies that the color symmetry 
of the strong interaction does not break spontaneously.\par
Using the algebraic rules 
in Eqs.(\ref{2.4})-(\ref{2.9}) 
and the shifting rule invoked in Ref.\ \cite{FSTM},
${\cal A}(x,y)$ can be rewritten as
\begin{equation}
 {\cal A}(x,y)=A_\mu(x,y)dx^\mu+{\mit\Phi}(x,y)\chi+G_\mu(x)dx^\mu, 
 \label{2.11}
\end{equation}
where
\begin{eqnarray}
&&    A_\mu(x,y) = \sum_{i}a_{i}^\dagger(x,y)\partial_\mu a_{i}(x,y), 
                                   \label{2.12}\\
&&     {\mit\Phi}(x,y) = \sum_{i}a_{i}^\dagger(x,y)\,(-a_i(x,y)M(y) 
            + M(y)a_i(x,-y))  \nonumber\\
&&\hskip 1.1cm  =a_{i}^\dagger(x,y)\,\partial_ya_i(x,y), \label{2.13}\\
&&  G_\mu(x)=\sum_{j}b_{j}^\dagger(x)\partial_\mu b_{j}(x).
  \label{2.14}
\end{eqnarray}
The functions $A_\mu(x,y)$, ${\mit\Phi}(x,y)$ and $G_\mu(x)$ here
are identified with the gauge field in the flavor symmetry, Higgs fields,
and the color gauge field responsible for the strong interaction,
respectively. 
The gauge transformations of these fields are well defined 
in the usual manner and  it follows that 
\begin{equation}
H(x,y)={\mit\Phi}(x,y)+M(y) \label{2.24a}
\end{equation}
is an un-shifted Higgs field whereas ${\mit\Phi}(x,y)$ denotes 
a shifted Higgs field with
 vanishing vacuum expectation value.
The nilpotency of ${\boldsymbol d}$ is proved 
using Eqs.(\ref{2.4})-(\ref{2.9}) along with
another algebraic rule in Ref.\ \cite{FSTM}.
\par
With these considerations, we can construct the gauge covariant field
strength,
\begin{equation}
  {\cal F}(x,y)= F(x,y) +  {\cal G}(x),
\label{2.26}
\end{equation}
where $F(x,y)$ and ${\cal G}(x)$ are the field strengths 
of flavor and color gauge fields, respectively, and given as
\begin{eqnarray}
&&     F(x,y) = {\boldsymbol d}A(x,y)+A(x,y)\wedge A(x,y),
     \nonumber\\
&&     {\cal G}(x)   =d\,G(x)+g_s G(x)\wedge G(x).
\label{2.27}
\end{eqnarray}
Using the algebraic rules defined in Eqs.(\ref{2.4})-(\ref{2.9}), we have
\begin{eqnarray}
 F(x,y ) &=& { 1 \over 2}F_{\mu\nu}(x,y)dx^\mu \wedge dx^\nu  \nonumber\\
           &&  + D_\mu {\mit\Phi}(x,y)dx^\mu \wedge \chi 
               + V(x,y)\chi \wedge \chi,
                \label{2.29}
\end{eqnarray}
where
\begin{eqnarray}
 && F_{\mu\nu}(x,y)=\partial_\mu A_\nu (x,y) - \partial_\nu A_\mu (x,y) 
               \nonumber\\
               &&\hskip 3cm +[A_\mu(x,y), A_\mu(x,y)],\label{2.30}\\
 && D_\mu {\mit\Phi}(x,y)=\partial_\mu {\mit\Phi}(x,y)
    + A_\mu(x,y)(M(y) + {\mit\Phi}(x,y))\nonumber\\
             && \hskip 2.5cm        -({\mit\Phi}(x,y)
                +M(y))A_\mu(x,-y),\label{2.31}\\
&&  V(x,y)= ({\mit\Phi}(x,y) + M(y))({\mit\Phi}(x,-y) \nonumber\\
       &&\hskip3cm        + M(-y)) - Y(x,y). \label{2.32}
\end{eqnarray}
The quantity 
$Y(x,y)$ in Eq.(\ref{2.32}) is auxiliary field and expressed as 
\begin{equation}
  Y(x,y)= \sum_{i}a_{i}^\dagger(x,y)M(y)M(-y)a_{i}(x,y).
 \label{2.33}
\end{equation}
This function  may become a constant field in the present construction.
In contrast to $F(x,y)$, ${\cal G}(x)$ is simply denoted as
\begin{eqnarray}   
 {\cal G}(x)&=&{1\over 2}{G}_{\mu\nu}(x)dx^\mu\wedge dx^\nu \nonumber\\
        &=&{1\over 2}\{\partial_\mu G^{}_\nu(x)-\partial_\nu G^{}_\mu(x)
        \nonumber\\
    &&\hskip 0.5cm   
     + g_s[G^{}_\mu(x), G^{}_\mu(x)]\}dx^\mu\wedge dx^\nu. 
\label{2.34}
\end{eqnarray}
\par
With the same metric structure as in Ref.\ \cite{FSTM}
 we can obtain the gauge invariant 
Yang-Mills-Higgs Lagrangian (YMH)
\begin{eqnarray}
{\cal L}_{\hbox{\rm\tiny{YMH}}}(x)&
=&-{\rm Tr}\sum_{y=\pm}{1 \over {\tilde g}^2}
< {\cal F}(x,y),  {\cal F}(x,y)>\nonumber\\
&=&-{\rm Tr}\sum_{y=\pm}{1\over 2{\tilde g}^2}
F_{\mu\nu}^{\dag}(x,y)F^{\mu\nu}(x,y)\nonumber\\
&&+{\rm Tr}\sum_{y=\pm}{1\over {\tilde g}^2}
    (D_\mu {\mit\Phi}(x,y))^{\dag}D^\mu {\mit\Phi}(x,y)  \nonumber\\
&& -{\rm Tr}\sum_{y=\pm}{1\over {\tilde g}^2}
        V^{\dag}(x,y)V(x,y)  \nonumber\\
&&-{\rm Tr}\sum_{y=\pm}{1\over 2{\tilde g}^2}{ G}_{\mu\nu}^{\dag}(x)
{ G}^{\mu\nu}(x).
\label{2.36}
\end{eqnarray}
Here, 
Tr denotes the trace over internal symmetry matrices including the color, 
flavor symmetries and generation space. 
The third term on the right-hand side 
is the potential term of the Higgs particle.
\par
\section{ An numerical estimation of the Higgs boson mass}
In reconstructing the standard model in the present scheme,
three generations of left and right-handed quarks and leptons
together with the  strong interaction must be taken into account.
A characteristic point of this formulation is to take the left 
and right-handed fermions $\psi(x,y)$ with arguments $x$ 
and $y(=\pm) $ in $M_4$ and $Z_2$, respectively, as
\begin{equation}
       \psi(x,+)=\left(\matrix{ 
                                u^r_{\hbox{\rm\tiny{L}}}\cr
                                u^g_{\hbox{\rm\tiny{L}}}\cr
                                u^b_{\hbox{\rm\tiny{L}}}\cr
                                \nu_{\hbox{\rm\tiny{L}}}\cr
                                d^r_{\hbox{\rm\tiny{L}}}\cr
                                d^g_{\hbox{\rm\tiny{L}}}\cr
                                d^b_{\hbox{\rm\tiny{L}}}\cr
                                e_{\hbox{\rm\tiny{L}}}\cr }
                            \right), 
\hskip 1.0cm
       \psi(x,-)=\left(\matrix{ 
                                u^r_{\hbox{\rm\tiny{R}}}\cr
                                u^g_{\hbox{\rm\tiny{R}}}\cr
                                u^b_{\hbox{\rm\tiny{R}}}\cr
                                 0       \cr
                                d^r_{\hbox{\rm\tiny{R}}}\cr
                                d^g_{\hbox{\rm\tiny{R}}}\cr
                                d^b_{\hbox{\rm\tiny{R}}}\cr
                                e_{\hbox{\rm\tiny{R}}}\cr }
                            \right), \label{2.1}
\end{equation}
where the subscripts L and R denote the left-handed and
right-handed fermions and the
superscripts $r$, $g$ and $b$ represent the color indices. 
It should be noted that $\psi(x,y)$ has an index for the three generations,
as do the explicit expressions for fermions on the right-hand sides
of Eq.$\,$(\ref{2.1}). 
Thus, $\psi(x,\pm)$ is a vector in the 24-dimensional space.
In order to construct the Dirac Lagrangian of the standard model
corresponding to $\psi(x,\pm)$ in Eqs.(\ref{2.1}), 
we  need a 24-dimensional generalized 
 covariant derivative composed of gauge and
Higgs fields on $M_4\times Z_2$. 
The gauge fields $A_\mu(x,y)$ and $G_\mu(x)$
in this covariant derivative must constitute
 the differential representation of 
the fermion fields in Eqs.(\ref{2.1}), and  therefore
they are expressed in $24\times24$ matrix forms. 
The Higgs field ${\mit\Phi}(x,y)$ is also taken 
to give the correct Yukawa interaction in the Dirac Lagrangian
and is expressed as a $24\times24$ matrix ( see Ref.\ \cite{FSTM}
for details).
We  find Yang-Mills-Higgs Lagrangian for the standard model 
as follows:
\begin{eqnarray}
{\cal L}_{\hbox{\rm\tiny{YMH}}}&=&
   -\frac14\sum_{k=1}^3\left(F_{\mu\nu}^k\right)^2 
   -\frac14B_{\mu\nu}^2 \nonumber\\
  &&  +|D_\mu h|^2   -\lambda(h^\dagger h-{\mu}^2)^2 \nonumber\\
 && - \frac{1}{4}
      \sum_{a=1}^8{G^a_{\mu\nu}}^{\dagger}{G^a}^{\mu\nu}, \label{3.20}
\end{eqnarray}
where
\begin{eqnarray}
   &&   F_{\mu\nu}^k=\partial_\mu A_\nu^k-\partial_\nu A_\mu^k
         +g\epsilon^{klm}A_\mu^lA_\nu^m,  \label{3.21}\\
   &&   B_{\mu\nu}=\partial_\mu B_\nu-\partial_\nu B_\mu,\label{3.22}\\
   &&     D^\mu h=[\,\partial_\mu-{i\over 2}\,(\sum_k\tau^kg{A^k_{L}}_\mu
          + \,\tau^0\,g'B_\mu\,)\,]\,h, \nonumber\\
     &&     \hskip 0cm h=\left(\matrix{ \phi^+ \cr
                                      \phi_0+\mu  \cr } \right),  
                                             \label{3.23}\\
   &&   G_{\mu\nu}^a=\partial_\mu G_\nu^a-\partial_\nu G_\mu^a
         +g_cf^{abc}G_\mu^bG_\nu^c, \label{3.24} 
\end{eqnarray}
with the following restrictions for coupling constants:
\begin{eqnarray}
  &&     g^2=\frac{{\tilde g}^2}{12}, \quad
      {g'}^2=\frac{{\tilde g}^2}{20}, \quad
      \lambda=\frac{{\tilde g}^2}{48}, \quad
        g_c^2=\frac{g_s^2{\tilde g}^2}{12}. \label{3.25}
\end{eqnarray}
Equation (\ref{3.25}) leads to the relation between coupling constants:
\begin{equation}
  g^2=\frac53{g'}^2=4\lambda\ne g_c^2,     \label{3.25a}
\end{equation}
which implies that the weak, electromagnetic and Higgs quartic coupling
constants become equal
and also
yields $\sin^2\theta_{\hbox{\rm\tiny W}}=3/8$ for 
the Weinberg angle as well as the  mass relations
\begin{eqnarray}
 &&       m_{\hbox{\rm\tiny W}}^2=\frac1{4}g^2v^2, \quad
          m_{\hbox{\rm\tiny Z}}^2=\frac{2}{5}g^2v^2, \quad
          m_{\hbox{\rm\tiny H}}^2= \frac{1}{2}g^2v^2.    
        \label{3.32}
\end{eqnarray}
with the vacuum expectation value of the Higgs boson $v$.
These relations hold only at tree level. Here, we assume
that these relations hold at a renormalization point and consider
their quantum effects  by use of the renormalization
group (RG) equations.
\par
With the notation 
\begin{equation}
   \alpha_3=\frac{g_c^2}{4\pi}, \quad
   \alpha_2=\frac{g^2}{4\pi}, \quad 
   \alpha_1={\frac53}\frac{{g'}^2}{4\pi},\quad
   \alpha_{\hbox{\rm\tiny H}}=\frac\lambda{4\pi}
   \label{4.2}
\end{equation}
for SU(3)$_c$, SU(2)$_{\hbox{\rm\tiny L}}$ 
and U(1)$_{\hbox{\rm\tiny Y}}$ gauges 
and the Higgs quartic coupling constants,
respectively,
the RG equations for these coupling constants are expressed as 
\begin{equation}
      \mu\frac{\partial\alpha_i}{\partial\mu}=\beta_i, 
      \: (i=1,2,3), \quad
    \mu\frac{\partial\alpha_{\hbox{\rm\tiny H}}}
    {\partial\mu}=\beta_{\hbox{\rm\tiny H}},
    \label{4.3}
\end{equation}
where the $\beta$-functions in the right hand sides are 
given in Ref.\ \cite{AMV} in the two loop approximation.
The Yukawa coupling constants of quarks written in $3\times3$ matrix form
in three generations are included in these $\beta$-functions.
We now assume the top quark mass  is dominant
in the evaluation of the RG equations. 
Masses of all particles in the standard model are introduced through
the vacuum expectation value $v$ of the Higgs field.
In this context, 
the running masses of gauge and Higgs bosons are defined as
\begin{equation}
    m_{\hbox{\rm\tiny W}}=\sqrt{\pi\alpha_2}v,\quad 
    m_{\hbox{\rm\tiny H}}
    =\sqrt{8\pi\alpha_{\hbox{\rm\tiny H}}}v.  \label{4.18}
\end{equation}
The top quark mass is also  expressed  as
\begin{equation}
    m_{\rm top}=\sqrt{2\pi\alpha_{\hbox{\rm\tiny Y}}}v,\label{4.19}
\end{equation}
with the Yukawa coupling constant $\alpha_{\hbox{\rm\tiny Y}}$ whose
RG equation is also given in Ref.\ \cite{AMV}.
The RG equation for $v$ is also given in Ref.\ \cite{AMV} 
in the two loop approximation.
\par
The RG equations for $\alpha_i\;(i=1,2,3)$, 
$\alpha_{\hbox{\rm\tiny H}}$, $\alpha_{\hbox{\rm\tiny Y}}$
and $v$ are highly non-linear equations with complicated coupling.
In order to solve these equations, we need six conditions.
Four of these conditions, those for $\alpha_i\;(i=1,2,3)$ and $v$, are 
given experimentally as \cite{AMV}
\begin{eqnarray}
  &&   \alpha_1(0)=0.017, \quad
     \alpha_2(0)=0.034,
     \quad
       \alpha_3(0)=0.12, \nonumber\\
  &&   v(0)=246{\rm GeV}, \label{4.24a}
\end{eqnarray}
at $\mu=m_{\hbox{\rm\tiny Z}}$ 
with the variable $t=\log(\mu/m_{\hbox{\rm\tiny Z}})$. 
That is, these conditions are given
at the neutral gauge boson mass $m_{\hbox{\rm\tiny Z}}=91.187$GeV.
According to Eq.(\ref{4.19}), the physical top quark mass $m_{\rm top}$ 
satisfies the equation
\begin{equation}
   m_{\rm top}=\sqrt{2\pi \alpha_{\hbox{\rm\tiny Y}}
   (t_{\rm top})}v(t_{\rm top}), \label{4.24b}
\end{equation}
where $t_{\rm top}=\log(m_{\rm top}/m_{\hbox{\rm\tiny Z}})$.
This equation constitutes one condition to solve the RG equations.
The remaining condition is
\begin{equation}
   \alpha_2(t_0)= \alpha_1(t_0)= 4\alpha_{\hbox{\rm\tiny H}}(t_0), 
   \label{4.24c}
\end{equation}
where $t_0$ is a constant, fixed in the numerical calculations.
The value of $t_0$ determines the energy 
at which the weak, electromagnetic and
 Higgs quartic interactions are unified.
With these considerations, we can find the running Higgs boson mass  
from Eq.(\ref{4.18}) as
\begin{equation}
          m_{\hbox{\rm\tiny H}}(t)
          =\sqrt{8\pi\alpha_{\hbox{\rm\tiny H}}(t)}v(t).  
          \label{4.27}
\end{equation}
The physical Higgs boson mass $m_{\hbox{\rm\tiny H}}$ is determined 
by imposing the condition that
\begin{equation}
        m_{\hbox{\rm\tiny H}}
        =\sqrt{8\pi\alpha_{\hbox{\rm\tiny H}}
        (t_{\hbox{\rm\tiny H}})}v(t_{\hbox{\rm\tiny H}}) 
         \label{4.28}
\end{equation}
with $t_{\hbox{\rm\tiny H}}
=\log(m_{\hbox{\rm\tiny H}}/m_{\hbox{\rm\tiny Z}})$.
\par
The top quark mass has a considerable effect on
the Higgs boson mass through Eq.(\ref{4.24b}). 
The top quark mass $m_{\rm top}$  reported in a review regarding work on
the top quark  \cite{PDG} is given as
\begin{equation}
     m_{\rm top}=175 \pm 6 {\rm GeV}. \label{4.29}
\end{equation}
We investigated the Higgs boson mass by varying the top quark mass 
in the range of Eq.(\ref{4.29}). Table 1. shows the physical
Higgs boson mass versus the top quark mass. 
Compared with the  one loop analysis of the Higgs boson mass 
in Ref.\ \cite{RGNCG}, two loop effects lower the numerical values 
by around 6GeV because, for example, 
$m_{\hbox{\rm\tiny H}}=164.01$GeV 
for $ m_{\rm top}=175$GeV in the one loop analysis.
This difference seems crucial since a Higgs boson with mass below 160GeV
could not decay into ${\rm W}^+{\rm W}^-$ 
and also experiments designed to search for the Higgs boson 
 depend greatly upon the present analysis.
It should be noted that owing to the unitarity requirement, 
$m_{\rm top}$ cannot exceed 190GeV because $\lambda$ would become  
minus at a much higher value of $t$ in such a case.
\par
It is interesting to investigate the running of the gauge and
Higgs quartic coupling constants, because these coupling constants
are unified at a point $t_0$ as shown in Eq.(\ref{4.24c}).
FIG. 1 displays the running of the three coupling constants.
\par
\vskip 0.5cm
\begin{minipage}[b]{8cm}
\leftline{Table 1.  Higgs boson mass
 versus top quark mass.}
\vskip 0.3cm
\begin{center}
\begin{tabular}{|c|c|}
\hline
         top quark (GeV) & Higgs boson (GeV) \\
\hline
         169.34   &      150.25   \\
\hline
         170.39  &       151.50   \\
\hline
         171.74   &      153.40  \\
\hline
         172.68   &      154.76 \\
\hline
         173.63 &        156.12 \\
\hline
         174.34   &      157.14  \\
\hline
         175.04  &       158.18  \\
\hline
         176.44  &       160.24\\
\hline
         177.37  &       161.61 \\
\hline
         178.29   &      163.00 \\
\hline
         179.65    &     165.07 \\
\hline
         180.34    &     166.12 \\
\hline
         181.02    &     167.15 \\
\hline
\end{tabular}
\end{center}
\end{minipage}
\hskip -0.5cm
\begin{minipage}[]{8cm}
 \begin{center}
    \epsfig{file=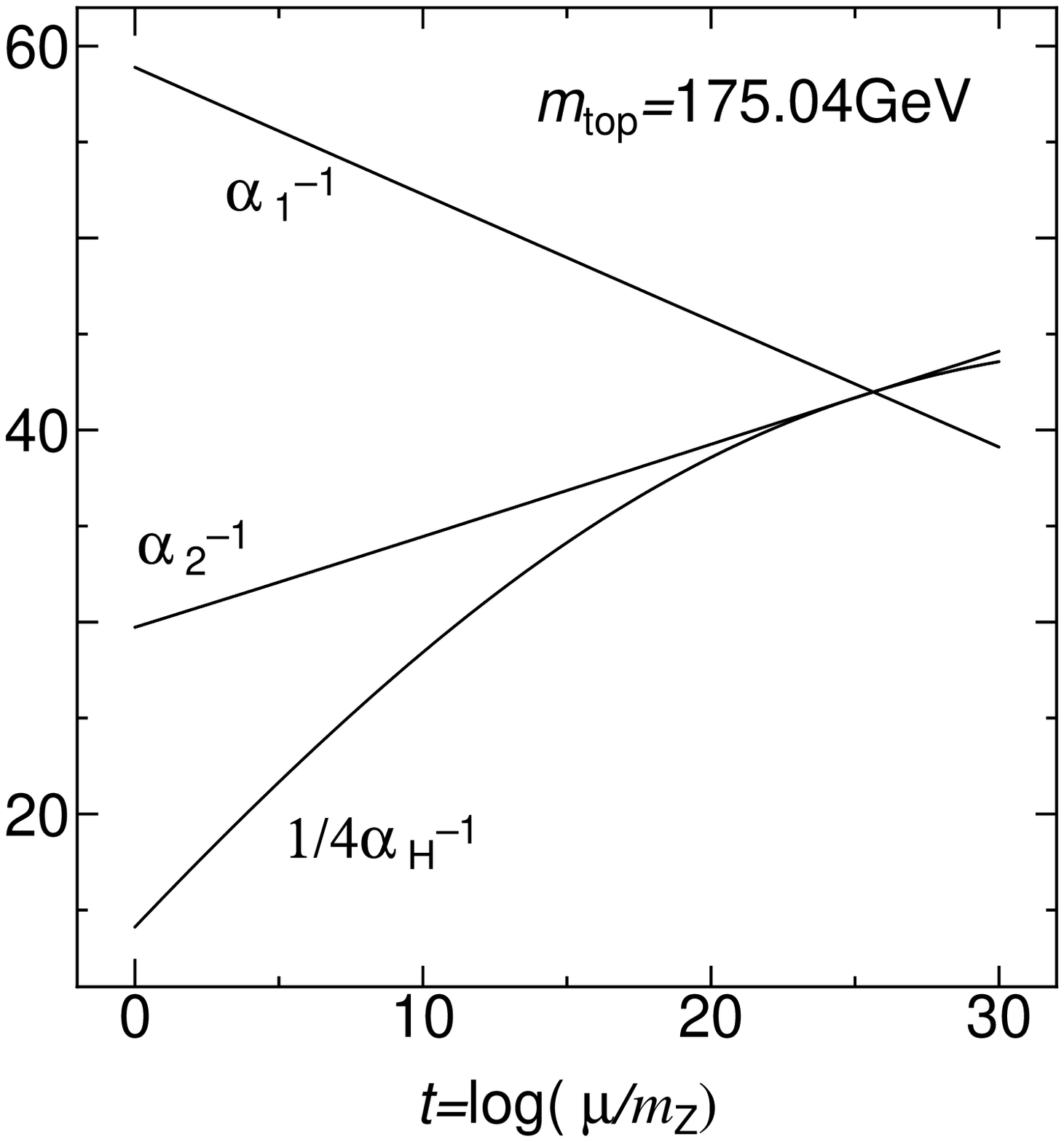,height=8cm, width=8cm}
  \vskip 0.0cm
Fig. 1: The running of the three coupling constants.
\end{center}
\end{minipage}
\par
\section{Conclusion}
In this paper,  we have 
  determined the Higgs boson mass by numerically solving the renomalization
  group equations with the relation 
  between coupling constants Eq.(\ref{3.25a}).
  It is introduced in the reconstruction of the standard model
  based on our new scheme of NCG \cite{FSTM}.
  We assumed that Eq.(\ref{3.25a}) holds at a renormalization point
  $t_0$. However, this leads to an interesting result that 
    the weak, electromagnetic and Higgs quartic coupling
  constants become equal at $t_0$ as shown Fig.1. In the case of
  $t_{\hbox{\rm\scriptsize top}}=175.04$, $t_0=26.635$ which amounts to 
  $\mu=3.37\times10^{13}$GeV. 
\par
We obtain $150.25\leq m_{\hbox{\rm\tiny H}} \leq 167.15$ in the
range of top quark mass $169.34\leq m_{\hbox{\rm\scriptsize top}}\leq 181.02$.
We hope that this result will be useful for experiments 
searching for the Higgs boson.


\end{document}